\documentclass[a4paper, fontsize=11pt, oneside, twocolumn=false, openright,
parskip=half-, titlepage=false, cleardoublepage=empty,
bibliography=totoc, BCOR=5mm, DIV=12]{scrartcl}
\usepackage{verbatim,csquotes}
\usepackage{graphicx,color,mathrsfs}
\usepackage{amsmath,amssymb}
\usepackage{xifthen} 
\usepackage{authblk}

\newcommand{\abs}[1]{\left| #1 \right|}
\newcommand{\br}[1]{\left( #1 \right)}
\newcommand{\brr}[1]{\left[ #1 \right]}

\newcommand{\e}{\mathrm{e}} 
\newcommand{\I}{\mathrm{i}} 

\newcommand{\ket}[1]{\vert {#1}\rangle}
\newcommand{\braket}[2]{\langle{#1} \vert {#2}\rangle}
\newcommand{\ketbra}[2]{\vert{#1} \rangle \langle{#2}\vert}
\newcommand{\HR}{H_\mathrm{R}}

\newcommand{\wc}{\omega_\mathrm{c}}
\newcommand{\F}{\mathcal{F}}
\newcommand{\psitar}{\psi_\mathrm{target}}
\newcommand{\psifin}{\psi_\mathrm{final}}
\newcommand{\psiin}{\psi_\mathrm{initial}}

\title{Optimized state transfer in systems of ultrastrongly coupled matter and radiation}

\author[1,2]{Luigi Giannelli}
\author[2,3]{Jishnu Rajendran}
\author[2,3]{Nicola Macrì}
\author[4,5,6]{Giuliano Benenti}
\author[7,8]{Simone Montangero}
\author[1,2,3]{Elisabetta Paladino}
\author[1,2,3]{Giuseppe Falci}

\affil[1]{CNR-IMM, UoS Università, 95123 Catania, Italy}
  
\affil[2]{Dipartimento di Fisica e Astronomia ``Ettore Majorana'', Universitá di
  Catania, 95123 Catania, Italy}

\affil[3]{INFN Sezione di Catania, 95123 Catania, Italy}

\affil[4]{Center for Nonlinear and Complex Systems, Dipartimento di Scienza e
  Alta Tecnologia, Università degli Studi dell’Insubria, 22100 Como, Italy}

\affil[5]{INFN Sezione di Milano, 20133 Milano, Italy}
  
\affil[6]{NEST, Istituto Nanoscienze-CNR, 56126 Pisa, Italy}

\affil[7]{Dipartimento di Fisica e Astronomia ``G. Galilei'', Università degli
  Studi di Padova, 35131 Padova, Italy}

\affil[8]{INFN, Sezione di Padova, 35131 Padova, Italy}

\date{}

\begin{document}

\maketitle

\begin{abstract}
  Ultrastrong coupling may allow faster operations for the development of
  quantum technologies at the expenses of increased sensitivity to new kind of
  intrinsic errors. We study state transfer in superconducting circuit QED
  architectures in the ultrastrong coupling regime. Using optimal control
  methods we find a protocol resilient to the main source of errors, coming from
  the interplay of the dynamical Casimir effect with cavity losses.
\end{abstract}

\section{Introduction: circuit-QED and adiabatic state transfer}
Circuit-QED solid-state systems~\cite{WallraffN2004strong} made of artificial
atoms (AA) and resonator modes~\cite{Haroche2006exploring} are paradigm models
for studying fundamental physics from measurement
theory~\cite{WallraffN2004strong} to quantum
thermodynamics~\cite{DiStefanoPRB2018nonequilibrium} and quantum
communication~\cite{BenentiPRL2009enhancement} besides being one of the most
promising platforms for quantum hardware~\cite{SchoelkopfN2008wiring}.
 
Recently solid-state ultrastrongly coupled (USC) AA-cavity systems have been
fabricated~\cite{Forn-DiazRMP2019ultrastrong,FriskKockumNRP2019ultrastrong}
where the coupling constants $g_i$ are comparable to the natural fequencies of
the AAs ($\epsilon^i$) and of the cavity ($\omega_c$). These structures may in
principle implement ultrafast quantum operations. The USC regime exhibits new
physical effects of great fundamental interest but detrimental for quantum
processing as the highly entangled nature of the eigenstates dressed by virtual
photons~\cite{Forn-DiazRMP2019ultrastrong,FriskKockumNRP2019ultrastrong,FalciSR2019ultrastrong}
and the triggering of photon pairs by the dynamical Casimir effect (DCE) when
coupling constants are time dependent. Multiphoton effects deteriorate the
fidelity of quantum operations~\cite{BenentiPRA2014dynamical} in USC
architectures even in absence of decoherence.

\subsection{Basic equations}
To overcome this problem a communication channel implemented by an adiabatic
protocol similar to STIRAP~\cite{VitanovRMP2017stimulated} has been
proposed~\cite{StramacchiaP2019speedup}. The system of two qubits (eigenstates
$\ket{\sigma^i} \in \{\ket{g^i},\ket{e^i}\}$ for $i=1,2$), coupled to a single
resonator mode is described by the Rabi Hamiltonian~\cite{Haroche2006exploring}
\begin{equation}
  \label{eq:RabiHamiltonian}
  \HR = \wc a^\dag a + \sum_{i=1,2} \epsilon^i \sigma_+^i\sigma^i_-
  + \sum_{i=1,2} g_i \br{a + a^\dag} \br{\sigma^i_- + \sigma^i_+},
\end{equation}
where $a$ ($a^\dag$) is the annihilation (creation) operator of the resonator
mode satisfying $[a,a^\dag]=1$, $\sigma^i_+ = \ketbra{e^i}{g^i}$ and $\sigma^i_-
= \ketbra{g^i}{e^i}$ are the qubit rising and lowering operators. We consider
resonant subsystems, $\wc=\epsilon^1=\epsilon^2$. The Hilbert space is spanned
by the factorized basis $\{\ket{n\sigma^2\sigma^1}\}$, where $\ket{n}$ are the
oscillator's number eigenstates.

The Rabi Hamiltonian is adopted when the couplings $g_i$ are large enough to
overcome the decoherence rates of the qubits ($\gamma^i$) and of the cavity
($\kappa$). If in addition $g_i \ll \wc$ the Hamiltonian conserves approximately
the number of excitation as described by the rotating wave approximation (RWA).
In the ultrastrong USC regime, where $0.1\wc\lesssim g_i\lesssim\wc$, this is no
longer true and the full Rabi Hamiltonian $\HR$ has to be taken into account.

We consider time-dependent couplings $g_i(t)$. In the regime where the RWA holds
a STIRAP-like process implemented by turning on $g_2(t)$ is before $g_1(t)$:
yields the state transfer~\cite{StramacchiaP2019speedup}
$\ket{0}\ket{g^2}\ket{\alpha g^1 + \beta e^1} = \ket{\psiin} \to
\ket{0}\ket{\alpha g^2 + \beta e^2}\ket{g^1} = \ket{\psitar}$. We seek the
performance of this protocol in the USC regime, allowing in principle much
faster operations. The figure of merit is the transfer efficiency
\begin{equation}
  \label{eq:fidelity}
  \F = \abs{\braket{\psitar}{\psifin}}^2,
\end{equation}
where $\ket{\psifin}$ is the state of the system at the end of the process
carried with the Hamiltonian Eq.~(\ref{eq:RabiHamiltonian}). As in
ref.~\cite{StramacchiaP2019speedup}, $\ket{\psifin}$ is obtained by solving the
Sch\"odinger equation $\I\partial_t\ket{\psi(t)} = \br{\HR(t) - \frac{\I}{2}
  \kappa a^\dag a}\ket{\psi(t)}$ with $\ket{\psi(-\infty)} = \ket{\psiin}$, the
extra non-Hermitian term describing cavity losses.

\begin{figure}[t]
  \centering
  \includegraphics{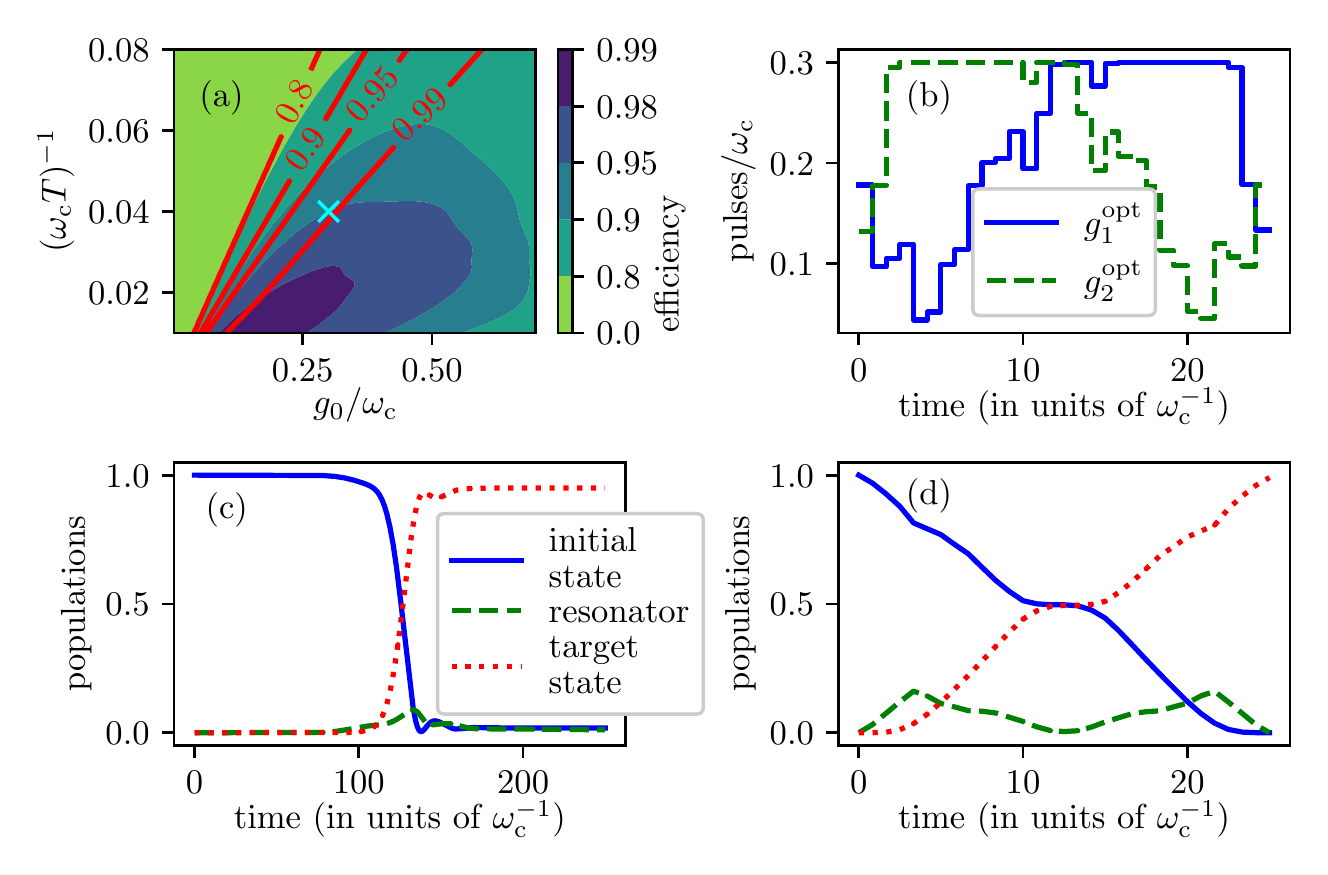}
  \caption{(a) Transfer efficiency $\F$ for Gaussian pulses $g_1(t) =
    g_0\e^{-\brr{\br{t-\tau}/T}^2}$ and $g_2(t) =
    g_0\e^{-\brr{\br{t+\tau}/T}^2}$ and $\kappa=0.005\wc$. The red lines
    represent constant $\F$ lines in the RWA where ideal STIRAP is implemented.
    The point $(\wc T)^{-1} = 0.04$, $g_0/\wc=0.3$ (cyan x) is analyzed in
    detail in the other panels: (b) shows the optimal shape of the couplings
    found by QOC; (c) shows the population histories for Gaussian pulses
    ($\F\simeq0.95$); (d) shows the population histories obtained with QOC
    ($\F\simeq0.99$). Notice the different time scales in panels (c) and (d)
    showing that the QOC protocol is faster by a factor of $\sim 3$.}
  \label{fig:evolution}
\end{figure}

Fig.~\ref{fig:evolution}(a) reports the transfer efficiency as a function of the
inverse speed $(\wc T)^{-1}$ and the maximal coupling $g_0/\wc$ for Gaussian
time dependence. We assume a cavity decay $\kappa=0.005 \wc$, which is a large
figure, compensating the oversimplified description of decoherence
sources~\cite{PaladinoNJP2011decoherence,PaladinoPRB2010optimal,PaladinoRMP2014oneoverfnoise}.
The red lines refer to what could be obtained if terms non conserving the number
of excitations were dropped, i.e. the RWA. For the Rabi model the efficiency is
still remarkably large even in the presence of cavity losses up to values $g
\sim 0.3\,\wc$. Here operations are already much faster than for standard
circuit-QED architectures where $g \sim 10^{-2}\,\wc$ and the RWA is applicable.
The efficiency is smaller than with the RWA in what could suggesting that an
important loss mechanism are virtual and DCE photons appearing during the
protocol, which are irreversibly lost by cavity decay.

\section{Results by optimal control}
In order to improve on the results of~\cite{StramacchiaP2019speedup} in what
follows we investigate the optimal time-dependence of $g_i(t)$ using the Quantum
Optimal Control (QOC)~\cite{GlaserEPJD2015training,AcinNJP2018quantum} tools
developed in~\cite{GiannelliAQ2021tutorial}.

We consider the point parameters $(\wc T)^{-1} = 0.04$, $g_0/\wc=0.3$ of
Fig.~\ref{fig:evolution}(a). The optimized $g_i^\mathrm{opt}(t)$ are step
functions shown in Fig.~\ref{fig:evolution}(b). The bottom panels of
Fig.~\ref{fig:evolution} report the evolution of the populations for Gaussian
pulses of width $T$ (c) and for the QOC solution $g_i^\mathrm{opt}(t)$ (d). For
the former protocol the duration is $\sim 3 T$ and $\F\simeq0.95$ while the
optimized pulses $g_i^\mathrm{opt}$ allow for $\F\simeq0.99$ (even slightly
better than RWA) in a time interval $T$, thus being also $\sim 3$ times faster.
Notice that the optimized pulses (Fig.~\ref{fig:evolution}(c)) are
counter-intuitively ordered thus this process is still STIRAP-like, with pulse
shapes performing better. For the QOC case it is apparent that the mode is more
populated during the protocol (see Fig.~\ref{fig:evolution}(d)) but this happens
for a shorter time so the impact of losses is reduced.

\section{Conclusions}
We have shown that QOC techniques can improve state transfer in a two-qubit
circuit-QED architecture in the USC regime. We found optimized
$g_i^\mathrm{opt}(t)$ yielding larger efficiency with a time duration shorter
than for Gaussian pulses with the same maximal strength $g_0$. It is likely that
QOC, machine learning
techniques~\cite{GiannelliAQ2021tutorial,BrownNJP2021reinforcement} or
superadiabatic driving~\cite{GiannelliPRA2014threelevel} could further improve
the transfer efficiency of such operations exploiting also modulation of
detunings~\cite{FalciFP2017advances,DiStefanoPRB2015population,DiStefanoPRA2016coherent}.

\section{Acknowledgments}
This work was supported by the QuantERA grant SiUCs (Grant No.731473), and by
University of Catania, Piano Incentivi Ricerca di Ateneo 2020-22, progtto Q-ICT.

\bibliographystyle{plain}
\bibliography{references}

\begin{thebibliography}{10}

\bibitem{AcinNJP2018quantum}
Antonio Ac{\'i}n, Immanuel Bloch, Harry Buhrman, Tommaso Calarco, Christopher
  Eichler, Jens Eisert, Daniel Esteve, Nicolas Gisin, Steffen~J. Glaser, Fedor
  Jelezko, Stefan Kuhr, Maciej Lewenstein, Max~F. Riedel, Piet~O. Schmidt, Rob
  Thew, Andreas Wallraff, Ian Walmsley, and Frank~K. Wilhelm.
\newblock The quantum technologies roadmap: A {{European}} community view.
\newblock {\em New J. Phys.}, 20(8):080201, August 2018.

\bibitem{BenentiPRL2009enhancement}
Giuliano Benenti, Antonio D'Arrigo, and Giuseppe Falci.
\newblock Enhancement of {{Transmission Rates}} in {{Quantum Memory Channels}}
  with {{Damping}}.
\newblock {\em Phys. Rev. Lett.}, 103(2):020502, July 2009.

\bibitem{BenentiPRA2014dynamical}
Giuliano Benenti, Antonio D'Arrigo, Stefano Siccardi, and Giuliano Strini.
\newblock Dynamical {{Casimir}} effect in quantum-information processing.
\newblock {\em Phys. Rev. A}, 90(5):052313, November 2014.

\bibitem{BrownNJP2021reinforcement}
Jonathon Brown, Pierpaolo Sgroi, Luigi Giannelli, Gheorghe~Sorin Paraoanu,
  Elisabetta Paladino, Giuseppe Falci, Mauro Paternostro, and Alessandro
  Ferraro.
\newblock Reinforcement learning-enhanced protocols for coherent
  population-transfer in three-level quantum systems.
\newblock {\em New J. Phys.}, 23(9):093035, September 2021.

\bibitem{DiStefanoPRB2018nonequilibrium}
P.~G. Di~Stefano, J.~J. Alonso, E.~Lutz, G.~Falci, and M.~Paternostro.
\newblock Nonequilibrium thermodynamics of continuously measured quantum
  systems: {{A}} circuit {{QED}} implementation.
\newblock {\em Phys. Rev. B}, 98(14):144514, October 2018.

\bibitem{DiStefanoPRB2015population}
P.~G. Di~Stefano, E.~Paladino, A.~D'Arrigo, and G.~Falci.
\newblock Population transfer in a {{Lambda}} system induced by detunings.
\newblock {\em Phys. Rev. B}, 91(22):224506, June 2015.

\bibitem{DiStefanoPRA2016coherent}
P.~G. Di~Stefano, E.~Paladino, T.~J. Pope, and G.~Falci.
\newblock Coherent manipulation of noise-protected superconducting artificial
  atoms in the {{Lambda}} scheme.
\newblock {\em Phys. Rev. A}, 93(5):051801, May 2016.

\bibitem{FalciFP2017advances}
G.~Falci, P.~G. Di~Stefano, A.~Ridolfo, A.~D'Arrigo, G.~S. Paraoanu, and
  E.~Paladino.
\newblock Advances in quantum control of three-level superconducting circuit
  architectures.
\newblock {\em Fortschritte der Physik}, 65(6-8):1600077, 2017.

\bibitem{FalciSR2019ultrastrong}
G.~Falci, A.~Ridolfo, P.~G. Di~Stefano, and E.~Paladino.
\newblock Ultrastrong coupling probed by {{Coherent Population Transfer}}.
\newblock {\em Sci Rep}, 9(1):9249, June 2019.

\bibitem{Forn-DiazRMP2019ultrastrong}
P.~{Forn-D{\'i}az}, L.~Lamata, E.~Rico, J.~Kono, and E.~Solano.
\newblock Ultrastrong coupling regimes of light-matter interaction.
\newblock {\em Rev. Mod. Phys.}, 91(2):025005, June 2019.

\bibitem{FriskKockumNRP2019ultrastrong}
Anton Frisk~Kockum, Adam Miranowicz, Simone De~Liberato, Salvatore Savasta, and
  Franco Nori.
\newblock Ultrastrong coupling between light and matter.
\newblock {\em Nat Rev Phys}, 1(1):19--40, January 2019.

\bibitem{GiannelliPRA2014threelevel}
Luigi Giannelli and Ennio Arimondo.
\newblock Three-level superadiabatic quantum driving.
\newblock {\em Phys. Rev. A}, 89(3):033419, March 2014.

\bibitem{GiannelliAQ2021tutorial}
Luigi Giannelli, Pierpaolo Sgroi, Jonathon Brown, Gheorghe~Sorin Paraoanu,
  Mauro Paternostro, Elisabetta Paladino, and Giuseppe Falci.
\newblock A {{Tutorial}} on {{Optimal Control}} and {{Reinforcement Learning}}
  methods for {{Quantum Technologies}}.
\newblock {\em arXiv:2112.07453 [quant-ph]}, December 2021.

\bibitem{GlaserEPJD2015training}
Steffen~J. Glaser, Ugo Boscain, Tommaso Calarco, Christiane~P. Koch, Walter
  K{\"o}ckenberger, Ronnie Kosloff, Ilya Kuprov, Burkhard Luy, Sophie Schirmer,
  Thomas {Schulte-Herbr{\"u}ggen}, Dominique Sugny, and Frank~K. Wilhelm.
\newblock Training {{Schr\"odinger}}'s cat: Quantum optimal control:
  {{Strategic}} report on current status, visions and goals for research in
  {{Europe}}.
\newblock {\em Eur. Phys. J. D}, 69(12):279, December 2015.

\bibitem{Haroche2006exploring}
Serge Haroche and Jean-Michel Raimond.
\newblock {\em Exploring the {{Quantum}}: {{Atoms}}, {{Cavities}}, and
  {{Photons}}}.
\newblock Oxford {{Graduate Texts}}. {Oxford University Press}, {Oxford}, 2006.

\bibitem{PaladinoNJP2011decoherence}
E.~Paladino, A.~D'Arrigo, A.~Mastellone, and G.~Falci.
\newblock Decoherence times of universal two-qubit gates in the presence of
  broad-band noise.
\newblock {\em New J. Phys.}, 13(9):093037, September 2011.

\bibitem{PaladinoRMP2014oneoverfnoise}
E.~Paladino, Y.~M. Galperin, G.~Falci, and B.~L. Altshuler.
\newblock \$\textbackslash mathbsf\{1\}/\textbackslash mathbsfit\{f\}\$ noise:
  {{Implications}} for solid-state quantum information.
\newblock {\em Rev. Mod. Phys.}, 86(2):361--418, April 2014.

\bibitem{PaladinoPRB2010optimal}
E.~Paladino, A.~Mastellone, A.~D'Arrigo, and G.~Falci.
\newblock Optimal tuning of solid-state quantum gates: {{A}} universal
  two-qubit gate.
\newblock {\em Phys. Rev. B}, 81(5):052502, February 2010.

\bibitem{SchoelkopfN2008wiring}
R.~J. Schoelkopf and S.~M. Girvin.
\newblock Wiring up quantum systems.
\newblock {\em Nature}, 451(7179):664--669, February 2008.

\bibitem{StramacchiaP2019speedup}
Michele Stramacchia, Alessandro Ridolfo, Giuliano Benenti, Elisabetta Paladino,
  Francesco Pellegrino, Daniele Maccarrone, and Giuseppe Falci.
\newblock Speedup of {{Adiabatic Multiqubit State-Transfer}} by {{Ultrastrong
  Coupling}} of {{Matter}} and {{Radiation}}.
\newblock {\em Proceedings}, 12(1):35, July 2019.

\bibitem{VitanovRMP2017stimulated}
Nikolay~V Vitanov, Andon~A Rangelov, Bruce~W Shore, and Klaas Bergmann.
\newblock Stimulated {{Raman}} adiabatic passage in physics, chemistry, and
  beyond.
\newblock {\em Reviews of Modern Physics}, 89(1):015006, 2017.

\bibitem{WallraffN2004strong}
A.~Wallraff, D.~I. Schuster, A.~Blais, L.~Frunzio, R.-S. Huang, J.~Majer,
  S.~Kumar, S.~M. Girvin, and R.~J. Schoelkopf.
\newblock Strong coupling of a single photon to a superconducting qubit using
  circuit quantum electrodynamics.
\newblock {\em Nature}, 431(7005):162--167, September 2004.

\end{thebibliography}

\end{document}